\documentclass[aps,amsmath,amssymb,10pt,floatfix,twocolumn,
showpacs,amfonts,floatfix,superscriptaddress,longbibliography,reprint]{revtex4-2}

\pdfoutput=1
\usepackage{amsthm,amsfonts,bbm}
\usepackage{mathrsfs}
\usepackage{lipsum}

\usepackage{subcaption}
\usepackage{graphicx}
\usepackage{ragged2e}
\usepackage{xcolor}
\usepackage[colorlinks=true,citecolor=blue,urlcolor=blue,linkcolor=blue,hyperindex]{hyperref}

\usepackage{adjustbox}

\usepackage{tikz,tikz-3dplot,tikz-dependency}
\usetikzlibrary{quotes,angles,arrows,arrows.meta,positioning,calc,snakes,fadings,shadings,intersections,decorations.text}
\usepackage{overpic}

\usepackage{algorithm}
\usepackage{algpseudocode}
\usepackage{tabularx}
\usepackage{braket}
\usepackage{booktabs}
\DeclareUnicodeCharacter{FF08}{(}
\DeclareUnicodeCharacter{FF09}{)}
\usepackage{yfonts}
\usepackage{color}
\usepackage[normalem]{ulem}
\usepackage{bm}
\usepackage{bbm}
\usepackage{mathtools}
\usepackage{array}
\usepackage{placeins}
\usepackage{enumitem}
\usepackage{float}
\usepackage[compat=0.6]{yquant}
\useyquantlanguage{groups}
\usetikzlibrary{fit,quotes,matrix,positioning}
\usetikzlibrary{trees}
\usepackage{listings}
\pgfdeclareshape{yquant-multictrlshape}{
	\inheritsavedanchors[from=yquant-circle]
	\foreach \anc in {center, north, north east, east, south east, south, south west, west, north west} {
		\inheritanchor[from=yquant-circle]{\anc}
	}
	\inheritanchorborder[from=yquant-circle]
	\backgroundpath{
		 \pgfpathmoveto{\pgfqpoint{-0.707107\dimexpr\xradius\relax}{-0.707107\dimexpr\yradius\relax}}
		 \pgfpathlineto{\pgfqpoint{0.707107\dimexpr\xradius\relax}{0.707107\dimexpr\yradius\relax}}
		\pgfpathellipse{\pgfpointorigin}
		{\pgfqpoint{\xradius}{0pt}}
		{\pgfqpoint{0pt}{\yradius}}
	}
	\inheritclippath[from=yquant-circle]
}

\yquantset{multictrl/.style={every control/.style={shape=yquant-multictrlshape,radius=1.05mm}}}
\yquantset{plusctrl/.style={/yquant/every control/.style={/yquant/operators/every not}, every positive control/.style={}}}

\def\<{\langle}
\def\>{\rangle}

\DeclareMathAlphabet\mathbfcal{OMS}{cmsy}{b}{n}

\mathchardef\mhyphen="2D 

\usepackage{amsmath,bm}
\usepackage{natbib}
\usepackage{siunitx}
\usepackage{witharrows}
\usepackage[title]{appendix}

\begin{document}
\title{Variational quantum algorithm for generalized eigenvalue problems of non-Hermitian systems}
	
\author{Jiaxin Li}

\author{Zhaobing Fan}

\author{Hongmei Yao}
\email{Contact author: hongmeiyao@163.com}

\author{Chunlin Yang}
\affiliation{School of Mathematical Sciences, Harbin Engineering University, Harbin 150001, China}

\author{Shao-Ming Fei}
\email{Contact author: feishm@cnu.edu.cn}
\affiliation{School of Mathematical Sciences, Capital Normal University, Beijing 100048, China}

\author{
  Zi-Tong Zhou,\textsuperscript{3}\\ 
Meng-Han Dou
}

\author{
Teng-Yang Ma
}

\affiliation{Origin Quantum Computing Technology Company Limited, Hefei 230088, China}

\begin{abstract}
	Non-Hermitian generalized eigenvalue problems (GEPs) play a significant role in many practical applications, such as  mechanical engineering. Based on the generalized Schur decomposition, we propose a variational quantum algorithm for solving the GEPs in non-Hermitian systems. The algorithm transforms the generalized eigenvalue problem into a process of searching for unitary transformation matrices. We demonstrate a method for evaluating both the loss function and its gradients on near-term quantum devices. We validate numerically the algorithm's performance through simulations, and demonstrate its application to GEPs in ocean acoustics. The algorithm's robustness is further confirmed through noise simulations.
\end{abstract}

\maketitle

\section{Introduction}
Generalized eigenvalue problems (GEPs) are of fundamental importance in both mathematics and applied science. Various methods have been developed to solve GEPs in classical computation \cite{B,MolerSIAM1973,SleijpenBit1996}. A key issue is that the memory usage and the computational complexity explode with the increasing system's scale, making the problems challenging for classical computers to solve large-scale GEPs \cite{4}.
	
The rapid development of quantum computing technology provides a solution to the above challenge \cite{5,6,7,8}. As a highly promising approach for high-performance computing, quantum computing offers significant speed advantages over classical computing due to its capability to handle exponentially large Hilbert spaces. The quantum phase estimation (QPE) algorithm, originally developed to calculate molecular ground-state energies in quantum chemistry \cite{Aspuru-Guzik,Malley}, has been extended in fault-tolerant quantum computers to solve the GEPs \cite{Parker,Raiche}.

Meanwhile, the variational quantum eigensolver (VQE), the classical-quantum hybrid framework designed to estimate ground state energies of Hamiltonians on near-term quantum devices \cite{12,14,15}, has been further extended to the variational quantum generalized eigensolver (VQGE) for computing the minimal generalized eigenvalues of Hamiltonians \cite{16,Sato,17}.

Both VQE and VQGE belong to the family of variational quantum algorithms (VQAs) \cite{13}, which employ parameterized quantum circuits (PQCs), also known as ansatze, and work by tuning their parameters to optimize a loss function that is typically the expectation value of an observable. This process forms a hybrid quantum-classical loop, where the quantum computer evaluates the loss function, and a classical optimizer uses this information to update the circuit parameters. VQAs have been widely applied to diverse problems and have emerged as a leading approach for demonstrating quantum heuristic advantage in the near future \cite{Kandala,Higgott,Liu,Singular}. However, the existing quantum algorithms are only applicable to the GEPs in Hermitian systems, and cannot be directly extended to non-Hermitian systems.

Currently, quantum algorithms capable of evaluating generalized eigenvalues in non-Hermitian systems remain scarce. Considering the broad applications of non-Hermitian generalized eigenvalue problems in ocean acoustics, mechanical engineering and many other related fields, developing quantum algorithms specifically for non-Hermitian systems carries substantial theoretical and practical importance.
	
In this paper, we present a VQGE for solving the GEPs in non-Hermitian systems. The remainder of this paper is organized as follows. Section \ref{Theoretical framework} establishes the theoretical framework of VQGE and defines a loss function. Section \ref{computation} elaborates the method for computing the loss function and its gradients on near-term quantum devices.
Section \ref{section:3} details the VQGE algorithm and the complexity analysis. Section \ref{section:5} conducts numerical simulations. Finally, we draw conclusions in Section \ref{section:6}.

\section{Theoretical framework and loss function}\label{Theoretical framework}
Let $A$ and $B$ be two $N\times N$ matrices in $\mathbb{C}^{N\times N}$. The generalized eigenvalue equation is expressed as
\begin{equation}
A|\psi\rangle=\lambda B|\psi\rangle,
\end{equation}
where $\lambda$ is the generalized eigenvalue of the matrix pair $(A,B)$, and $|\psi\rangle$ is the corresponding eigenvector. Let $\lambda(A, B)$ denote the set of generalized eigenvalues of the matrix pair $(A, B)$.
	
To determine the generalized eigenvalues of $(A, B)$, our variational quantum generalized eigensolver employs the generalized Schur decomposition theory \cite{G}, which states that for any $A,B\in \mathbb{C}^{N\times N}$, there exist unitary matrices $Q$ and $Z$ such that
	\begin{equation}\label{x}
		Q^{\dag}AZ=T,~\ Q^{\dag}BZ=S,
	\end{equation}
where $T = (t_{ij})$ and $S = (s_{ij})$ for $i, j = 1, 2, \cdots, N$ are upper triangular matrices in $ \mathbb{C}^{N \times N}$. If $t_{kk} = s_{kk} = 0$ for some $k \in \{1, 2, \cdots, N \}$, then $\lambda(A, B)=\mathbb{C}$. Otherwise,
	\begin{equation*}
		\lambda(A, B)=\left\{\frac{t_{ii}}{s_{ii}}|\ s_{ii}\neq 0, i=1,2,\cdots, N \right\}.
	\end{equation*}
	
Based on the Schur decomposition, for given matrices $A,B\in \mathbb{C}^{2^n\times 2^n}$, we define a loss function
	\begin{equation}\label{loss}
		\begin{split}
			\mathcal{L}(\boldsymbol{\theta},\boldsymbol{\phi})= &\sum_{i=1}^{2^n-1}\sum_{j=0}^{i-1}\left(|\langle i|Q^{\dag}(\boldsymbol{\theta})AZ(\boldsymbol{\phi})|j\rangle|^{2} \right. \\
			&+ \left.|\langle i|Q^{\dag}(\boldsymbol{\theta})BZ(\boldsymbol{\phi})|j\rangle|^{2}\right) \\
			= &\sum_{i=1}^{2^n-1}\sum_{j=0}^{i-1}\left(|\langle i|T(\boldsymbol{\theta},\boldsymbol{\phi})|j\rangle|^{2} + |\langle i|S(\boldsymbol{\theta},\boldsymbol{\phi})|j\rangle|^{2}\right),
		\end{split}
	\end{equation}
where $T(\boldsymbol{\theta},\boldsymbol{\phi})=Q^{\dag}(\boldsymbol{\theta})AZ(\boldsymbol{\phi})$ and $S(\boldsymbol{\theta},\boldsymbol{\phi})=Q^{\dag}(\boldsymbol{\theta})BZ(\boldsymbol{\phi})$. The matrices $Q(\boldsymbol{\theta})$ and $Z(\boldsymbol{\phi})$ are unitary and parameterized by the vectors $\boldsymbol{\theta}=(\theta_1,\theta_2,\cdots,\theta_{\ell_1})$ and $\boldsymbol{\phi}=(\phi_1,\phi_2,\cdots,\phi_{\ell_2})$, respectively. The state $|i\rangle$ denotes the $i$-th computational basis.
	
\textbf{Theorem 1.} The loss function $\mathcal{L}(\boldsymbol{\theta},\boldsymbol{\phi})$ attains its global minimum of zero if and only if $T(\boldsymbol{\theta},\boldsymbol{\phi})$ and $S(\boldsymbol{\theta},\boldsymbol{\phi})$ are upper triangular matrices.
	
\paragraph*{Proof.} Since $\mathcal{L}(\boldsymbol{\theta},\boldsymbol{\phi})$ is a sum of non-negative terms, zero is its global minimum.

($\Rightarrow$) If $\mathcal{L}(\boldsymbol{\theta},\boldsymbol{\phi})=0$, then
$|\bra{i}T(\boldsymbol{\theta},\boldsymbol{\phi})\ket{j}|^2=|\bra{i}S(\boldsymbol{\theta},\boldsymbol{\phi})\ket{j}|^2=0$
for $i=1,\cdots, 2^n-1$ and $j=0,\cdots i-1$, since all the terms in (\ref{loss}) are non-negative. This implies that $T(\boldsymbol{\theta},\boldsymbol{\phi})$ and $S(\boldsymbol{\theta},\boldsymbol{\phi})$ must be upper triangular matrices.
	
($\Leftarrow$) If $T(\boldsymbol{\theta},\boldsymbol{\phi})$ and $S(\boldsymbol{\theta},\boldsymbol{\phi})$ are upper triangular, it is direct to verify that $\mathcal{L}(\boldsymbol{\theta},\boldsymbol{\phi})=0$.
	\hfill $\square$

\section{Computation of loss function and its gradient}\label{computation}

\subsection{Quantum implementation of loss function}

\begin{figure}[H]
	\centering
	\begin{tikzpicture}[scale=0.92]
		\begin{yquant}
			qubit {$\ket{0}$} w0;
			qubit {$\ket{0}$} w1;
			nobit  w2;
			qubit {$\ket{0}$} w3;
			qubit {$\ket{0}$} an0;
			qubit {$\ket{0}$} an1;
			nobit an2;
			qubit {$\ket{0}$} an3;
			qubit {$\ket{0}$} au0;
			qubit {$\ket{0}$} au1;
			nobit  au2;
			qubit {$\ket{0}$} au3;
			
			text {$\vdots$}  w2;
			text {$\vdots$}  au2;
			text {$\vdots$}  an2;
			
			hspace {5pt} -;
			box {$H$} w0;
			box {$H$} w1;
			box {$H$} w3;
			box {$H$} an0;
			hspace {5pt} -;
			cnot au0 | w0;
			hspace {5pt} -;
			cnot au1 | w1;
			hspace {5pt} -;
			text {$\cdots$}  w0;
			text {$\cdots$}  w1;
			text {$\cdots$}  w3;
			text {$\cdots$}  an0;
			text {$\cdots$}  an1;
			text {$\cdots$}  an3;
			text {$\cdots$}  au0;
			text {$\cdots$}  au1;
			text {$\cdots$}  au3;
			hspace {5pt} -;
			cnot au3 | w3;
			hspace {5pt} -;
			[x radius=0.5cm, y radius=0.5cm]
			box {$Z(\boldsymbol{\phi})$} (au0,au1,au2,au3);
			hspace {5pt} -;
			[x radius=0.5cm, y radius=0.5cm]
			box {$U_{A}$} (an1,an2,an3,au0,au1,au2,au3)~ an0;
			hspace {5pt} -;
			[x radius=0.5cm, y radius=0.5cm]
			box {$U_{B}$} (an1,an2,an3,au0,au1,au2,au3)| an0;
			hspace {5pt} -;
			[x radius=0.5cm, y radius=0.5cm]
			box {$Q^{\dagger}(\boldsymbol{\theta})$} (au0,au1,au2,au3);
			hspace {10pt} -;
			measure w0,w1,w3,an0,au0,au1,au3,an1,an3;
			text {$\vdots$}  w2;
			text {$\vdots$}  au2;
			text {$\vdots$}  an2;
		\end{yquant}
		\draw [decorate, decoration={brace, amplitude=6pt, mirror}, xshift=-7mm]
		(0,-0.1) -- (0,-2.4)
		node [midway, left=10pt] {w($n$)};
		\node at (-1.5, -2.9) {idx(1)};
		\draw [decorate, decoration={brace, amplitude=6pt, mirror}, xshift=-7mm]
		(0,-3.5) -- (0,-5)
		node [midway, left=10pt] {a($m$)};
		\draw [decorate, decoration={brace, amplitude=6pt, mirror}, xshift=-7mm]
		(0,-5.5) -- (0,-7.5)
		node [midway, left=10pt] {aug($n$)};
	\end{tikzpicture}
	\caption{Quantum circuit for computation of the loss function in (\ref{loss}).}\label{figer:2}
\end{figure}
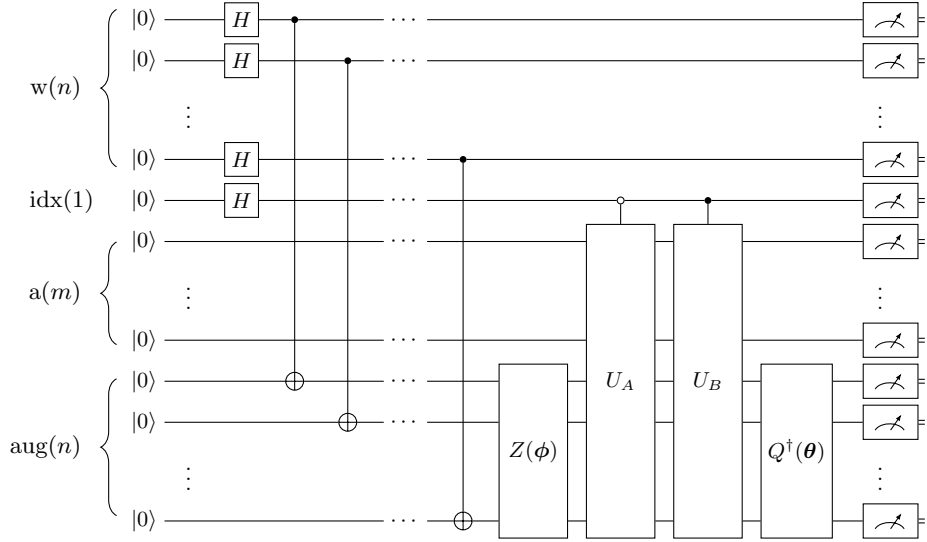

We employ the quantum process snapshot (QPS) technique \cite{14}, which measures the entries of a matrix in a single quantum circuit, to compute the loss function. For convenience, we incorporate index register to enable simultaneous measuring the entries of many matrices within a single quantum circuit, see the implementations in Appendix \ref{section:QPS}. The loss function $\mathcal{L}(\boldsymbol{\theta},\boldsymbol{\phi})$ can be computed via the quantum circuit shown in Fig. \ref{figer:2}, where $Q^{\dagger}(\boldsymbol{\theta})$ and $Z(\boldsymbol{\phi})$ are parameterized quantum circuits. When the matrices $A$ and $B$ are non-unitary, they can be encoded into quantum systems through data input models such as block encoding \cite{37,Yang, LZX} or linear combination of unitaries (LCU) \cite{38,Childs}. Unitary matrices $U_A$ and $U_B$ are employed to realize the quantum encoding of $A$ and $B$, respectively. In this work, we employ the LCU method as the foundational framework for quantum data input, see the quantum circuit in Appendix \ref{section:Data}. Without loss of generality, we decompose matrices $A$ and $B$ into linear combinations of $2^m$ unitary matrices. The registers follow this naming convention: a$(m)$ denotes ancilla register with $m$ qubits, w$(n)$ denotes work register with $n$ qubits, idx(1) denotes index register with 1 qubit, and aug$(n)$ denotes augmented register with $n$ qubits.

\onecolumngrid

\twocolumngrid

We denote CNOT$^n$ the $n$ CNOT gates with the control qubits in the work register and the target qubits in the augmented register, establishing a qubit-by-qubit correspondence. To implement matrices $A$ and $B$ on the augmented register, the measurement outcome of the ancilla register must be $|0\rangle^{\otimes m }$. When the state of the ancilla register collapses to $|0\rangle^{\otimes m}$, the quantum circuit evolves the initial state $\ket{0}^{\otimes n}_{\text{w}} \ket{0}_{\text{idx}} \ket{0}_{\text{aug}}^{\otimes n}$ as follows:

\begin{widetext}
\begin{equation}\label{evolve}
	\begin{split}	
		&\ket{0}^{\otimes n}_{\text{w}} \ket{0}_{\text{idx}} \ket{0}_{\text{aug}}^{\otimes n}\\
		\xrightarrow{H^{\otimes n}\otimes H \otimes I^{\otimes n}}&\frac{1}{\sqrt{2^{n+1}}}	 \sum_{i=0}^{2^n-1}
		\Big( \ket{i}_{\text{w}} \ket{0}_{\text{idx}} \ket{0}_{\text{aug}}^{\otimes n} 
		+  \ket{i}_{\text{w}} \ket{1}_{\text{idx}} \ket{0}_{\text{aug}}^{\otimes n} \Big)\\
		\xrightarrow{\text{CNOT}^{n}}&\frac{1}{\sqrt{2^{n+1}}}\sum_{i=0}^{2^n-1}	\Big( \ket{i}_{\text{w}} \ket{0}_{\text{idx}} \ket{i}_{\text{aug}}
		+  \ket{i}_{\text{w}} \ket{1}_{\text{idx}} \ket{i}_{\text{aug}}  \Big)\\
		\xrightarrow{I^{\otimes n}\otimes I \otimes Z(\boldsymbol{\phi})}&
		\frac{1}{\sqrt{2^{n+1}}}\sum_{i=0}^{2^n-1}\Big(  \ket{i}_{\text{w}} \ket{0}_{\text{idx}} Z(\boldsymbol{\phi}) \ket{i}_{\text{aug}}
		+  \ket{i}_{\text{w}} \ket{1}_{\text{idx}} Z(\boldsymbol{\phi}) \ket{i}_{\text{aug}} \Big)\\
		\xrightarrow{I^{\otimes n}\otimes \left( \ket{0}\bra{0} \otimes A + \ket{1}\bra{1}\otimes B \right) } &
		\frac{1}{\sqrt{2^{n+1}}}\sum_{i=0}^{2^n-1}
		\Big( \ket{i}_{\text{w}} \ket{0}_{\text{idx}} AZ(\boldsymbol{\phi}) \ket{i}_{\text{aug}}
		+  \ket{i}_{\text{w}} \ket{1}_{\text{idx}} BZ(\boldsymbol{\phi}) \ket{i}_{\text{aug}} \Big)\\
		\xrightarrow{I^{\otimes n}\otimes I \otimes Q^{\dagger}(\boldsymbol{\theta})}
		&\frac{1}{\sqrt{2^{n+1}}}\sum_{i=0}^{2^n-1}
		\Big( \ket{i}_{\text{w}} \ket{0}_{\text{idx}} Q^{\dagger}(\boldsymbol{\theta})AZ(\boldsymbol{\phi}) \ket{i}_{\text{aug}} 
		+  \ket{i}_{\text{w}} \ket{1}_{\text{idx}} Q^{\dagger}(\boldsymbol{\theta})BZ(\boldsymbol{\phi}) \ket{i}_{\text{aug}} \Big)\\
		=&\frac{1}{\sqrt{2^{n+1}}}\sum_{i=0}^{2^n-1}
		\Big( \ket{i}_{\text{w}} \ket{0}_{\text{idx}} T(\boldsymbol{\theta},\boldsymbol{\phi}) \ket{i}_{\text{aug}}
		+  \ket{i}_{\text{w}} \ket{1}_{\text{idx}}S(\boldsymbol{\theta},\boldsymbol{\phi}) \ket{i}_{\text{aug}} \Big).
	\end{split}
\end{equation}

\end{widetext}

\textbf{Theorem 2.} Let $P_L$ denote the probability of measurement results in the set $L=\{|j\rangle_{\text{w}}|0\rangle_{\text{idx}}\ket{0}_{\text{a}}^{\otimes m}|i\rangle_{\text{aug}},\ |j\rangle_{\text{w}}|1\rangle_{\text{idx}}\ket{0}_{\text{a}}^{\otimes m}|i\rangle_{\text{aug}}\mid i=1,2,\cdots,2^n-1,\ j=0,1,\cdots,i-1\}$. The loss function is then computed as
\begin{equation}\label{C}
	\mathcal{L}(\boldsymbol{\theta},\boldsymbol{\phi})=2^{n+1}P_L.
\end{equation}

\paragraph*{Proof.} According to (\ref{evolve}), the probability of observing a state $|j\rangle_{\text{w}}|k\rangle_{\text{idx}}\ket{0}_{\text{a}}^{\otimes m}|i\rangle_{\text{aug}}$ is
\begin{equation}\label{6}
	P_{ijk} =
	\begin{cases}
		\frac{1}{2^{n+1}} |\bra{i}T(\boldsymbol{\theta},\boldsymbol{\phi})\ket{j}|^2,   &k=0,\\
		\frac{1}{2^{n+1}} |\bra{i}S(\boldsymbol{\theta},\boldsymbol{\phi})\ket{j}|^2,   &k=1.
	\end{cases}
\end{equation}
Summing these probabilities over all states in the set $L$, we obtain the total probability,
\begin{equation*}
	\begin{split}
	P_{L} &= \frac{1}{2^{n+1}}\sum_{i=1}^{2^n-1}\sum_{j=0}^{i-1}\left(|\langle i|T(\boldsymbol{\theta},\boldsymbol{\phi})|j\rangle|^{2} + |\langle i|S(\boldsymbol{\theta},\boldsymbol{\phi})|j\rangle|^{2}\right)\\
	&=\frac{1}{2^{n+1}} \mathcal{L}(\boldsymbol{\theta},\boldsymbol{\phi}).
\end{split}
\end{equation*}
Thus, we obtain
\begin{equation*}
\mathcal{L}(\boldsymbol{\theta},\boldsymbol{\phi})= 2^{n+1} P_L.
\end{equation*}
\hfill $\square$

Theorem 2 reveals the relationship between the loss function and the probabilities of specific measurement outcomes. In practical implementations, our goal is to estimate each term $T_{ij}=|\langle i|T(\boldsymbol{\theta},\boldsymbol{\phi})|j\rangle|^{2}$ and $S_{ij}=|\langle i|S(\boldsymbol{\theta},\boldsymbol{\phi})|j\rangle|^{2}$ for $i=1,2,\cdots,2^n-1$ and $j=0,1,\cdots,i-1$. Each term is estimated from the frequency of its corresponding quantum state observed in a finite number of measurements.

We denote the total number of measurements by $N_{\text{meas}}$.
When the bit string $\ket{i}_{\text{w}}\ket{0}_{\text{idx}}\ket{0}_{\text{a}}^{\otimes m}\ket{j}_{\text{aug}} $ corresponding to $T_{ij}$ is observed $f_{ij}^{(T)}$ times, and the bit string $\ket{i}_{\text{w}}\ket{1}_{\text{idx}}\ket{0}_{\text{a}}^{\otimes m}\ket{j}_{\text{aug}} $ corresponding to $S_{ij}$ is observed $f_{ij}^{(S)}$ times, the experimental estimates $\tilde{T}_{ij}$ and $\tilde{S}_{ij}$ for each term are given by (\ref{6}) as
\begin{equation}
\frac{1}{2^{n+1}}\tilde{T}_{ij}=\frac{f_{ij}^{(T)}}{N_{\text{meas}}},\  \frac{1}{2^{n+1}}\tilde{S}_{ij}=\frac{f_{ij}^{(S)}}{N_{\text{meas}}}.
\end{equation}
This process can be viewed as a Bernoulli trial. After repeating the experiment $N_{\text{meas}}$ times, we aim to bound the probability that the relative error of our estimates exceeds a predefined threshold $c>0$. The relative errors for the squared magnitudes are defined as
\begin{equation}
\varepsilon_{ij}^{(T)}=\frac{|\tilde{T}_{ij}-T_{ij}|}{T_{ij}},\ 
\varepsilon_{ij}^{(S)}=\frac{|\tilde{S}_{ij}-S_{ij}|}{S_{ij}}.
\end{equation}
Applying Hoeffding's inequality \cite{35}, for any $c >0$, we obtain
\begin{equation} \label{Pr}
\begin{split}
\Pr(\varepsilon_{ij}^{(T)}\geqslant c)&\leqslant \text{e} ^{-2N_{\text{meas}}c^{2}{(\frac{1}{2^{n+1}}T_{ij})}^2 },\\
\Pr(\varepsilon_{ij}^{(S)}\geqslant c)&\leqslant \text{e} ^{-2N_{\text{meas}}c^{2}{(\frac{1}{2^{n+1}}S_{ij})}^2 }.
\end{split}
\end{equation}
The equation above shows that the probability of a large relative error decreases exponentially with the number of measurements $N_{\text{meas}}$. To ensure that the error probability for each term remains below a given $\epsilon$, we require
\begin{equation}
	\begin{split}
		& \text{e} ^{-2N_{\text{meas}}c^{2}{(\frac{1}{2^{n+1}}T_{ij})}^2 }\leqslant \epsilon,\\
		&\text{e} ^{-2N_{\text{meas}}c^{2}{(\frac{1}{2^{n+1}}S_{ij})}^2 }\leqslant \epsilon.
	\end{split}
\end{equation}
Solving for $N_{\text{meas}}$ for each inequality, we find for all $i=1,2,\cdots,2^n-1$ and $j=0,1,\cdots,i-1$, 
\begin{equation}\label{bound}
N_{\text{meas}}\geqslant \max\limits_{i,j}\left\{2^{2n+1}\frac{\ln(1/\epsilon)}{c^2T_{ij}^2}, \  2^{2n+1}\frac{\ln(1/\epsilon)}{c^2S_{ij}^2}\right\}.
\end{equation}
The sampling cost $N_{\text{meas}}$ is explicitly given by the lower bound in (\ref{bound}). This cost depends on the number of qubits $n$ and the smallest $T_{ij}$ and $S_{ij}$ during optimization. However, strictly following this lower bound during the early stages of optimization is inefficient. The primary goal at this initial phase is to find the correct direction to minimize the loss function, rather than to measure every element with high precision. Therefore, we adopt an adaptive strategy that begins with a smaller number of measurements and progressively increases this number as the loss function decreases, thereby ensuring accurate results upon convergence. As indicated by (\ref{bound}), the sampling cost scales exponentially with the number of qubits $n$, implying rapidly growing resource requirements for large systems. While this exponential scaling presents a challenge for large-scale problems, the measurement cost remains manageable for problem sizes relevant to near-term quantum devices, ensuring the algorithm's practicality in the near term.

\subsection{Gradient estimation with the parameter shift rule}
The essential goal of variational quantum-classical hybrid algorithms is to find the optimal parameters $\boldsymbol{\theta}_{opt}$ and $\boldsymbol{\phi}_{opt}$. While both gradient-based and gradient-free optimization methods are available, this work employs the gradient-based parameter shift rule \cite{Mitarai}, a widely adopted approach in variational quantum circuits. It outperforms the finite difference method \cite{Nocedal},  which estimates the partial derivative using $\frac{\partial}{\partial \theta_{i}}f(\boldsymbol{\theta})\approx \frac{f(\boldsymbol{\theta}+\Delta \theta_{i})-f(\boldsymbol{\theta})}{\Delta \theta_{i}}$, where $\Delta \theta_{i}$ must be sufficiently small for the approximation to hold. The finite difference method is highly sensitive to noise. This sensitivity arises from both the amplification of precision errors when $\Delta \theta_{i}$ is too small and the inherent shot noise in quantum measurements. The sensitivity to shot noise can be explicitly quantified using statistical inequalities. Let $g=\frac{f(\boldsymbol{\theta}+\Delta \theta_{i})-f(\boldsymbol{\theta})}{\Delta \theta_{i}}$ be the target value to estimate. Consider its experimental estimator $\tilde{g}=\frac{\tilde{f}(\boldsymbol{\theta}+\Delta \theta_{i})-\tilde{f}(\boldsymbol{\theta})}{\Delta \theta_{i}}$, where $\tilde{f}$ is the finite-sample estimate of the expectation value $f$ obtained from $N_{\text{meas}}$ measurements on a quantum device. For a bounded observable, $\tilde{f}$ satisfies $|\tilde{f}| \leqslant B$. According to Hoeffding's inequality, the probability that the estimation error exceeds a tolerance $\tau>0$ satisfies
\begin{equation}\label{finitedifference}
	\Pr(|\tilde{g}-g|\geqslant \tau)\leqslant 2 \text{e}^{-\frac{N_{\text{meas}}{(\tau \Delta \theta_{i})}^2}{8B^2}}.
\end{equation}
As shown in (\ref{finitedifference}), keeping the probability that the error exceeds $\tau$ below a small constant $\nu$ requires 
\begin{equation}
	2 \text{e}^{-\frac{N_{\text{meas}}{(\tau \Delta \theta_{i})}^2}{8B^2}}\leqslant\nu.
\end{equation}
Solving this inequality for $N_{\text{meas}}$ yields
\begin{equation} \label{14}
N_{\text{meas}}\geqslant \frac{8B^2}{{(\tau \Delta \theta_i)}^2} \ln(\frac{2}{\nu}).
\end{equation}
Eq. (\ref{14}) shows that $N_{\text{meas}}$ must grow as $\mathcal{O}(1/{(\Delta \theta_{i})}^2)$. This quadratic scaling makes high-precision gradient estimation via the finite difference method unaffordably expensive on noisy devices. In contrast, with the parameter shift rule, the gradient can be estimated directly and is less sensitive to perturbations as long as the gradient is non-vanishing.

In the algorithm implementation, we apply quantum gate sequences $Q=Q_{\ell_1}\cdots Q_{1}$ and $Z=Z_{\ell_2}\cdots Z_{1}$ to perform unitary transformations on matrices $A$ and $B$. Each gate $Q_l$ and $Z_k$ is either fixed, e.g., CNOT gate, or parameterized, for all $l=1,\cdots, \ell_1$ and $k=1,\cdots,\ell_2$. The parameterized gates $Q_l$ and $Z_k$ take the form $Q_l=\text{e}^{-iH_l\theta_l/2}$ and $Z_k=\text{e}^{-iV_k\phi_k/2}$, where $\theta_l$ and $\phi_k$ are real parameters, and $H_l$ and $V_k$ are tensor products of Pauli matrices. This choice of gate generators allows us to compute the gradient of the loss function $\mathcal{L}(\boldsymbol{\theta},\boldsymbol{\phi})$ on near-term quantum devices using the parameter shift rule \cite{Mitarai}.
The gradient vector is given by
\begin{equation}
	\nabla \mathcal{L}(\boldsymbol{\theta},\boldsymbol{\phi})=(\frac{\partial \mathcal{L}}{\partial \theta_1},\cdots,\frac{\partial \mathcal{L}}{\partial \theta_{\ell_1}},
	\frac{\partial \mathcal{L}}{\partial \phi_1},\cdots,\frac{\partial \mathcal{L}}{\partial \phi_{\ell_2}} ).
\end{equation}
We compute each partial derivative via
\begin{equation}\label{derivatives}
	\begin{split}
		\frac{\partial \mathcal{L}(\boldsymbol{\theta},\boldsymbol{\phi})}{\partial \theta_l}&=\frac{\mathcal{L}(\boldsymbol{\theta}_{l+\frac{\pi}{2}},\boldsymbol{\phi})-
			\mathcal{L}(\boldsymbol{\theta}_{l-\frac{\pi}{2}},\boldsymbol{\phi})}{2},\\
		\frac{\partial \mathcal{L}(\boldsymbol{\theta},\boldsymbol{\phi})}{\partial \phi_k}&=\frac{\mathcal{L}(\boldsymbol{\theta},\boldsymbol{\phi}_{k+\frac{\pi}{2}})-
			\mathcal{L}(\boldsymbol{\theta},\boldsymbol{\phi}_{k-\frac{\pi}{2}})}{2},
	\end{split}
\end{equation}
where $\boldsymbol{\theta}_{l\pm \frac{\pi}{2}}=(\theta_1,\cdots, \theta_{l}\pm \frac{\pi}{2}, \cdots, \theta_{\ell_1}) $ and $\boldsymbol{\phi}_{k\pm \frac{\pi}{2}}=(\phi_1,\cdots, \phi_{k}\pm \frac{\pi}{2}, \cdots, \phi_{\ell_2}) $ denote parameter vectors with shifts applied only to the $l$-th or $k$-th component, respectively. Appendix \ref{proof} establishes the applicability of the parameter shift rule given in (\ref{derivatives}) to our loss function.

\section{Variational quantum generalized eigensolver}\label{section:3}
Our VQGE for non-Hermitian GEPs is presented in Algorithm \ref{algorithm:1}. The inputs to the algorithm include the LCU circuits $U_A$ and $U_B$ for matrices $A$ and $B$, as well as the parameterized circuits $Q^{\dag}(\boldsymbol{\theta})$ and $Z(\boldsymbol{\phi})$. The algorithm then enters a hybrid quantum-classical optimization loop to minimize the loss function $\mathcal{L}(\boldsymbol{\theta},\boldsymbol{\phi})$ defined in (\ref{loss}). An error threshold $\varepsilon$ is set as the stopping criterion. Once $\mathcal{L}(\boldsymbol{\theta},\boldsymbol{\phi})<\varepsilon$ is met, the optimal parameters $\boldsymbol{\theta}_{opt}$ and $\boldsymbol{\phi}_{opt}$ are obtained. These are used to construct the matrices $T=Q^{\dagger}(\boldsymbol{\theta}_{opt})AZ(\boldsymbol{\phi}_{opt})$ and $S=Q^{\dagger}(\boldsymbol{\theta}_{opt})BZ(\boldsymbol{\phi}_{opt})$. The generalized eigenvalues are estimated from the diagonal elements of $T$ and $S$.

\begin{algorithm}[H]
	\caption{Variational quantum generalized eigensolver (VQGE)}\label{algorithm:1}
	\begin{algorithmic}[1]
		\State $\textbf{Input:}$ $U_A$, $U_B$, parametrized circuits $Q^{\dag}(\boldsymbol{\theta})$ and $Z(\boldsymbol{\phi})$ with initial parameters of $\boldsymbol{\theta}_{0}$, $\boldsymbol{\phi}_{0}$, and error threshold \(\varepsilon\);
		\State  Compute the loss function \(\mathcal{L}(\boldsymbol{\theta},\boldsymbol{\phi})\) and its gradient;
		\While{$\mathcal{L}(\boldsymbol{\theta},\boldsymbol{\phi})$ has not converged}
		\State $\theta_{i} \gets \theta_{i} - \delta \frac{\partial \mathcal{L}}{\partial \theta_{i}}$;
		\State $\phi_{k} \gets \phi_{k} - \delta \frac{\partial \mathcal{L}}{\partial \phi_{k}}$;
		\EndWhile
		\State \Return $\boldsymbol{\theta}_{opt}, \boldsymbol{\phi}_{opt}$;
		\State  Let $T=Q^{\dagger}(\boldsymbol{\theta}_{opt})AZ(\boldsymbol{\phi}_{opt})$, $S=Q^{\dagger}(\boldsymbol{\theta}_{opt})BZ(\boldsymbol{\phi}_{opt})$;
		\State  Obtain the diagonal elements \(\{t_{ii}\}_{i=1}^{N}\) and \(\{s_{ii}\}_{i=1}^{N}\) of matrices \(T\) and \(S\) using Hadamard test;
		\State  Use the diagonal elements \(\{t_{ii}\}_{i=1}^{N}\) and \(\{s_{ii}\}_{i=1}^{N}\) to evaluate the generalized eigenvalues of the matrix pair \((A,B)\).
	\end{algorithmic}
\end{algorithm}

\textbf{Theorem 3.} Concerning the computational complexity of computing the loss function, the gate complexity is $\mathcal{O}(n+2^{m}\text{poly}(n))$ and the qubit complexity is $\mathcal{O}(2n+ m+1)$.

\paragraph*{Proof.}
The QPS technique requires only $\mathcal{O}(n)$ gates in the variational quantum algorithm. To implement non-unitary matrices $A$ and $B$ via LCU, they must be decomposed into $2^m$ unitary terms, requiring $\mathcal{O}(2^m\text{poly}(n))$ gates. Additionally, the ansatz circuits $Q^{\dag}(\boldsymbol{\theta})$ and $Z(\boldsymbol{\phi})$ each require $\mathcal{O}(n)$ gates. The total gate complexity of the quantum circuit is $\mathcal{O}(n+2^{m}\text{poly}(n))$.

As shown in Fig. \ref{figer:2}, the work register contains $n$ work qubits, the index register contains 1 qubit, the ancilla register contains $m$ qubit and the augmented register contains $n$ qubits. Thus, the total qubit complexity is $\mathcal{O}(2n+ m+1)$.
\hfill $\square$

Compared to the classical generalized Schur decomposition with $\mathcal{O}(2^{3n})$ computational complexity, when matrices $A$ and $B$ can be expressed as linear combinations of few unitaries ($n \gg m$, e.g., sparse or $k$-local Hamiltonians), the VQGE algorithm can demonstrate exponential quantum heuristic advantage. Therefore, in practice, minimizing the number of unitary matrices in the decomposition of $A$ and $B$ is critical for optimal performance.

It should be noted that the trainability of variational quantum algorithms can sometimes be hindered by the barren plateau phenomenon, where the gradient of the loss function vanishes exponentially with the number of qubits \cite{Jarrod}. However, this is not a universal issue. It mainly appears when using hardware-efficient ansatze and global loss functions \cite{Kandala,Singular}. In such cases, traditional optimizers often do not work well. To address this, current strategies include using identity-block initialization \cite{Edward}, employing local cost functions \cite{NatCommun12}, and constructing adaptive Hamiltonians \cite{35}. For our specific VQGE algorithm, the loss function $\mathcal{L}(\boldsymbol{\theta},\boldsymbol{\phi})$ has a local structure, as each term depends only on specific quantum basis states. This locality helps to reduce the likelihood of barren plateau problem.

\section{Numerical results}\label{section:5}

We simulate the VQGE on the Origin Quantum cloud platform with Pyqpanda \cite{36}. The results confirm the feasibility of our proposed algorithm.

\subsection{Two-qubit system} \label{section:A}

Consider the randomly generated two-qubit matrix pair $(A,B)$, where
\begin{equation*}
	\begin{split}
		A&=\begin{pmatrix}
			-0.846053 & -3.121318 & 1.130982 & -0.135525\\
			-0.274860 & 0.540084 & 0.832479 & 0.530499\\
			-0.135770 & 0.613640 & 0.947157 & -0.638468\\
			1.730607 & -1.242851 & -2.299600 & 0.060833
		\end{pmatrix},\\
		B&=\begin{pmatrix}
			0.217329 & 0.418199 & 1.206862 & 1.458747\\
			-0.208682 & -1.124809 & 0.288132 & 2.032686\\
			1.272089 & -0.145261 & 1.799622 & 1.183555\\
			0.000000 & 0.000000 & 0.000000 & 0.000000
		\end{pmatrix}.
	\end{split}
\end{equation*}

\begin{figure}[H]
	\centering
	\includegraphics[scale=0.6]{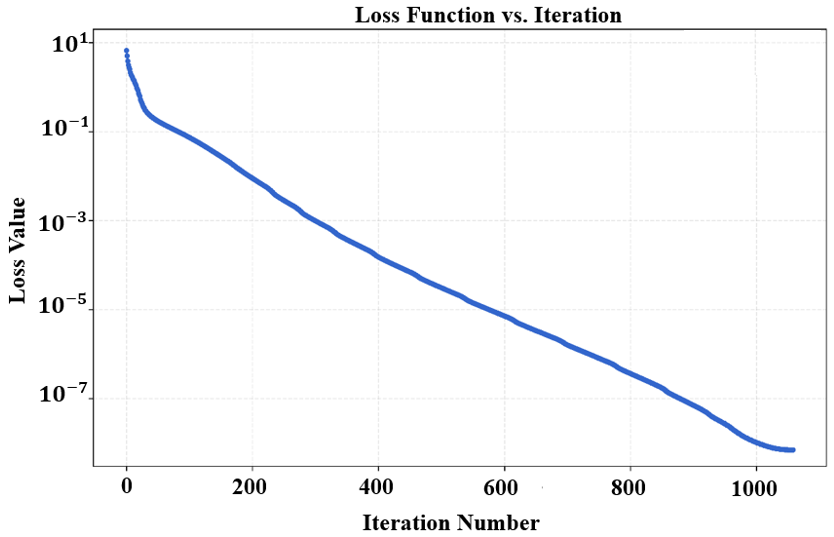}
	\caption{The iterative process for computing the generalized eigenvalues in two-qubit system.}\label{figer:4}
\end{figure}

As shown in Fig. \ref{figer:4}, the loss function converges to the order of $10^{-7}$ after 900 iterations. The parameters $\boldsymbol{\theta}$ and $\boldsymbol{\phi}$ used at the 900th iteration are denoted as $\boldsymbol{\theta}_{opt}$ and $\boldsymbol{\phi}_{opt}$, respectively. Table \ref{Table1} presents the exact values, experimental values, and percentage relative errors of the generalized eigenvalues for the matrix pair $(A,B)$. The first three experimental eigenvalues show excellent agreement with the exact values. Since rank$(B)=3$, the matrix $S$ contains one diagonal element approaching zero, which leads to an extremely large experimental value. According to the generalized Schur decomposition theory, the fourth experimental value should not be considered a valid generalized eigenvalue. This behavior perfectly matches the theoretical prediction, confirming the algorithm's effectiveness.

\begin{table}[H] 
		\caption{Comparison of exact and experimental generalized eigenvalues in two-qubit system.} \label{Table1} 
	\begin{tabular}{cccc} 
		\toprule 
		Index & Exact values & Experimental values & Error (\%) \\
		\midrule 
		1 & -4.650054 & -4.650453-0.000675i & 0.0086\% \\
		2 & 0.211286+0.223149i & 0.211293+0.223152i & 0.0025\% \\
		3 & 0.211286-0.223149i & 0.211283-0.223145i & 0.0015\% \\
		4 & / & -14755.94+27008.85i & / \\
		\bottomrule 
	\end{tabular}
\end{table}

\subsection{Application in ocean acoustic fields} \label{section:B}
\begin{figure}[H]
	\centering
	\includegraphics[scale=0.68]{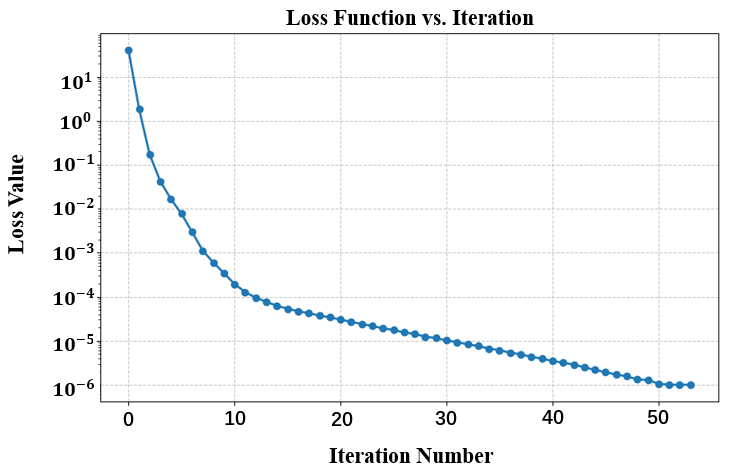}
	\caption{The iterative process for computing the generalized eigenvalues in ocean acoustic fields.}\label{figer:6}
\end{figure}

\begin{figure}[H]
	\centering
	\begin{subfigure}[b]{0.45\textwidth}
		\includegraphics[width=\textwidth]{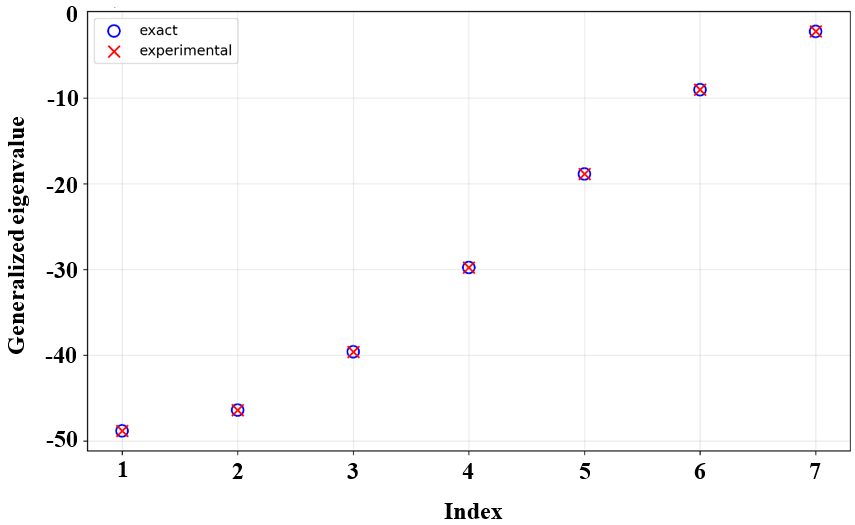}
		\caption{Comparison of the first 7 generalized eigenvalues.}
		\label{fig:sub1}
	\end{subfigure}
	\hfill
	\begin{subfigure}[b]{0.45\textwidth}
		\includegraphics[width=\textwidth]{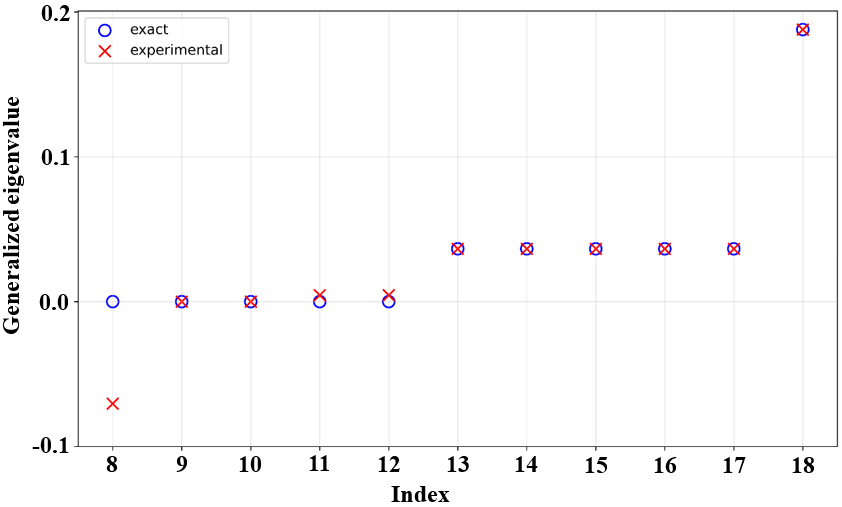}
		\caption{Comparison of the 8th to 18th generalized eigenvalues.}
		\label{fig:sub2}
	\end{subfigure}
	\caption{Comparison of exact and experimental generalized eigenvalues in ocean acoustic fields.}
	\label{fig:comparison}
\end{figure}

Ocean acoustics is fundamental to investigate underwater sound propagation characteristics. Based on the normal-mode theory for ocean acoustic propagation, we employ the governing equations for sound pressure modes in seawater and stress-displacement modes in ice-seawater environments. The sound pressure mode $\varphi_{m}\left(z\right)$ satisfies the Helmholtz equation \cite{jensen}
\begin{equation*}
\frac{d^2 \varphi_{m}\left(z\right)}{dz^2} + \left(\frac{\omega^{2}}{c^{2}\left(z\right)} - k_{m}^{2}\right)\varphi_{m}\left(z\right) = \boldsymbol{0},
\end{equation*}
where $z$ is the depth, $c(z)$ is the sound velocity, $k_m$ is the horizontal wave number and $\omega$ is the angle frequency of sound wave.  The stress-displacement mode $\xi_m(z)=(\xi_{m,1}(z),\xi_{m,2}(z),\xi_{m,3}(z),\xi_{m,4}(z))^{\rm T}$ follows the coupled differential equations
\begin{equation*}
	\begin{split}
		&\begin{cases}
			\begin{split}
				\frac{d\xi_{m,1}\left(z\right)}{dz} &= -\xi_{m,2}\left(z\right) + \frac{1}{\mu(z)} \xi_{m,4}\left(z\right), \\
				\frac{d\xi_{m,2}\left(z\right)}{dz} &= \frac{\lambda(z)k_m^2}{\lambda(z)+2\mu(z)} \xi_{m,1}\left(z\right) + \frac{1}{\lambda(z)+2\mu(z)} \xi_{m,4}\left(z\right), \\
				\frac{d\xi_{m,3}\left(z\right)}{dz} &= \left(\frac{4\mu(z)\left(\lambda(z)+\mu(z)\right)k_m^2}{\lambda(z)+2\mu(z)} - \rho(z)\omega^2\right) \xi_{m,1}\left(z\right) \\
				&\quad - \frac{\lambda(z)}{\lambda(z)+2\mu(z)}\xi_{m,4}\left(z\right), \\
				\frac{d\xi_{m,4}\left(z\right)}{dz} &= -\rho(z)\omega^2 \xi_{m,2}\left(z\right) + k_m^2 \xi_{m,3}\left(z\right).
			\end{split}
		\end{cases}
	\end{split}
\end{equation*}
where $\lambda(z)$ and $\mu(z)$ are the depth-dependent Lam\'{e} coefficients of ice. Based on finite difference method, the problem is converted into a generalized eigenvalue problem, $AV=BV\Sigma$,
where $A$ and $B$ are non-Hermitian matrices, $\Sigma$ is a diagonal matrix with the generalized eigenvalues on its diagonal, and $V$ is a matrix whose column vectors are the corresponding generalized eigenvectors \cite{jensen,aki}. After simplification, matrices $A$ and $B$ can be reduced to $32 \times 32$ non-Hermitian sparse matrices. Their specific structures and element values are provided in \cite{Yang,40}. Due to the singularity of the matrix $B$, we employ a projection method to map $B$ to its non-singular subspace for enhanced algorithmic efficiency \cite{G}. The generalized eigenvalues of the projected matrices remain highly consistent
with those of the original matrices. Fig. \ref{figer:6} shows the iterative convergence of the corresponding loss function, reaching stable convergence after 50 iterations. Therefore, we select the  parameters $\boldsymbol{\theta}$ and $\boldsymbol{\phi}$ obtained at the 50th iteration as $\boldsymbol{\theta}_{opt}$ and $\boldsymbol{\phi}_{opt}$ to estimate the generalized eigenvalues of the matrix pair $(A,B)$. In this practical study, only real generalized eigenvalues are required. When the imaginary part of the eigenvalue is sufficiently close to zero, it can be omitted. Fig. \ref{fig:comparison} shows the comparison between experimental results and exact values. Since $\text{rank}(B) = 18$, there are at most 18 generalized eigenvalues. The results demonstrate good agreement between experimental values and theoretical values.

\subsection{Noise-incorporated simulation} \label{section:C}

\begin{figure}[H]
	\centering
	\begin{minipage}[t]{0.46\textwidth}
		\centering
		\includegraphics[width=\linewidth]{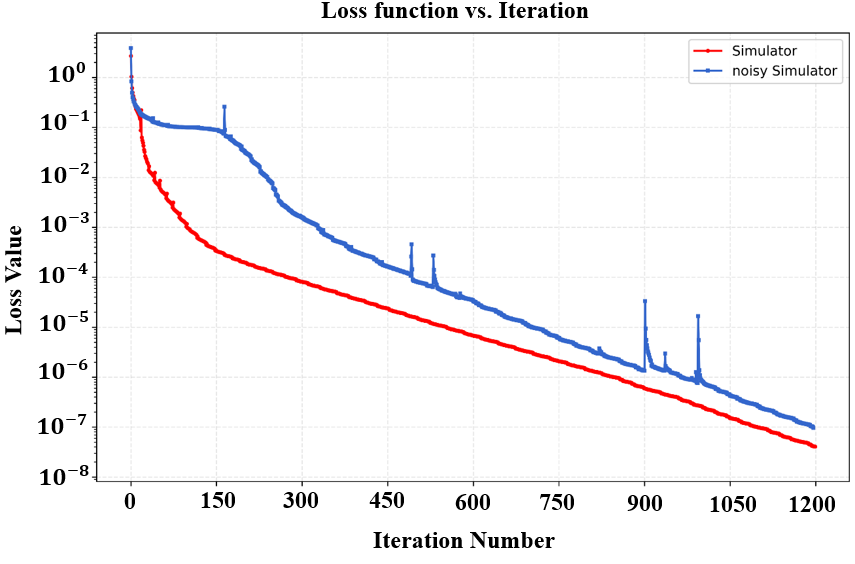}
		\captionof{figure}{The iterative process under noisy and noiseless conditions.}
		\label{figer:5}
	\end{minipage}
	\hfill
	\begin{adjustbox}{valign=t, raise=4cm}
		\begin{minipage}[t]{0.48\textwidth}
			\centering
			\captionof{table}{Comparison of evaluated and theoretical generalized eigenvalues under noisy condition.} 
			\begin{tabular}{cccc} 
				\toprule 
				Index & Exact values & Experimental values & Error (\%) \\
				\midrule 
				1 & 2.240765 & 2.239893+0.000698i & 0.0389\% \\
				2 & 0.750178 & 0.750185-0.000139i & 0.0009\% \\
				3 & 0.270846 & 0.270977-0.000100i & 0.0484\% \\
				4 & 1.428571 & 1.429336+0.000472i & 0.0536\% \\
				\bottomrule 
			\end{tabular}
			\label{Table2}
		\end{minipage}
	\end{adjustbox}
\end{figure}

Quantum noise presents a major challenge for implementing quantum algorithms. To thoroughly evaluate algorithm performance, we perform simulations using noise models in Pyqpanda. We tested randomly generated two-qubit matrix pairs $(A, B)$.
Fig. \ref{figer:5} compares the loss function iteration processes under noisy and noiseless conditions. The results show that despite noise-induced fluctuations, the loss function consistently decreases below $10^{-7}$ after 1200 iterations. We select the $\boldsymbol{\theta}$ and $\boldsymbol{\phi}$ from the 1200th iteration under noisy conditions as
$\boldsymbol{\theta}_{opt}$ and $\boldsymbol{\phi}_{opt}$. Table \ref{Table2} demonstrates good consistency between experimental and exact values. While our noise-incorporated simulations focus on a two-qubit system due to classical computational constraints, the results demonstrate the inherent robustness of VQGE under typical noise channels (e.g., amplitude damping and depolarizing noise). For larger systems, we expect the noise resilience to depend on the locality of the Hamiltonian and error mitigation strategies. Several error mitigation techniques, including zero-noise extrapolation, quasi-probability method, Clifford data regression, and quantum subspace expansion, can be used to improve algorithm performance on noisy devices \cite{Errormitigation,TemmeErrormitigation2017,EndoPractical2018,CzarnikError2021,McCleanHybrid2017}. Our next step will be to test the VQGE algorithm on real quantum hardware using these methods.

\onecolumngrid

\twocolumngrid


\section{Conclusion}\label{section:6}

We have proposed a variational quantum algorithm for solving the generalized eigenvalues problem in non-Hermitian systems by using the generalized Schur decomposition, designing a new loss function, and demonstrating how to compute both the loss function and its gradients on near-term quantum devices. We have validated the algorithm's performance through numerical simulations and shown its application to generalized eigenvalue problems in ocean acoustics. Additional noise simulations confirm the algorithm's robustness. Our results demonstrate the feasibility of computing the generalized eigenvalues for non-Hermitian systems on near-term quantum devices.

\section*{Acknowledgements}
This work was supported by the Stable Supporting Fund of Acoustic Science and Technology Laboratory (JCKYS2024604SSJS001), and the Fundamental Research Funds for the Central Universities (3072024XX2401).

\appendix

\section{Quantum process snapshot}\label{section:QPS}
\begin{figure}[H]
	\centering
	\begin{tikzpicture}[scale=1]
		\begin{yquant}
			qubit {work} l;
			qubit {aug} k;
			["north:$n$"
			{font=\protect\footnotesize, inner sep=0pt}]
			slash l;
			["north:$n$"
			{font=\protect\footnotesize, inner sep=0pt}]
			slash k;
			hspace {5pt} -;

			box {$H^{\otimes n}$} l;
			hspace {5pt} -;
			box {CNOT$^n$} (l,k);
			hspace {5pt} -;
			box {$U$} (k);
			hspace {5pt} -;
			measure l,k;
		\end{yquant}
	\end{tikzpicture}
\caption{Quantum circuit of QPS \cite{14}.}\label{QPS}
\end{figure}
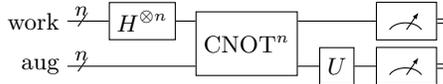

The QPS technique measures all squared moduli $\{|\langle i|U|j\rangle|^{2}|i,j=0,1,\cdots ,2^n-1\}$ of a $2^n\times 2^n$ unitary matrix $U$ by using the circuit shown in Fig. \ref{QPS} \cite{14}. The quantum circuit evolves the initial state $\ket{0}^{\otimes n}_{\text{w}} \ket{0}_{\text{aug}}^{\otimes n}$ as follows:
\begin{equation}
	\begin{split}
		&|0\rangle^{\otimes n}_{\text{w}}|0\rangle^{\otimes n}_{\text{aug}}\\
		\xrightarrow{H^{\otimes n} \otimes I^{\otimes n}}&\frac{1}{\sqrt{2^{n}}}\sum_{i=0}^{2^n-1}\ket{i}_{\text{w}}\ket{0}^{\otimes n}_{\text{aug}}\\
		\xrightarrow{ \text{CNOT}^n}&\frac{1}{\sqrt{2^{n}}}\sum_{i=0}^{2^n-1}\ket{i}_{\text{w}}\ket{i}_{\text{aug}}\\
		\xrightarrow{I^{\otimes n}\otimes U}&\frac{1}{\sqrt{2^{n}}}\sum_{i=0}^{2^n-1}\ket{i}_{\text{w}}U\ket{i}_{\text{aug}}.
	\end{split}
\end{equation}
Thus, the probability of obtaining the measurement outcome $|j\rangle_{\text{w}}|i\rangle_{\text{aug}}$ is $\frac{1}{2^n}|\langle i|U|j\rangle|^{2}$.

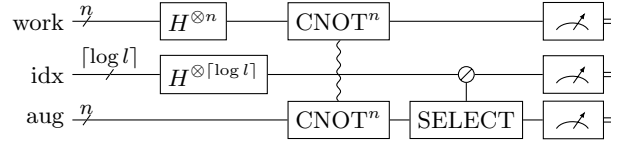
\begin{figure}[H]
	\centering
	\begin{tikzpicture}[scale=1]
		\begin{yquant}
			qubit {work} l;
			qubit {idx} j;
			qubit {aug} k;
			["north:$n$"
			{font=\protect\footnotesize, inner sep=0pt}]
			slash l;
			["north:$\lceil\log {l}\rceil$"
			{font=\protect\footnotesize, inner sep=0pt}]
			slash j;
			["north:$n$"
			{font=\protect\footnotesize, inner sep=0pt}]
			slash k;
			hspace {5pt} -;
			
			box {$H^{\otimes n}$} l;
			box {$H^{\otimes \lceil \log l\rceil}$} j;
			hspace {5pt} -;
			box {CNOT$^n$} (l,k);
			hspace {5pt} -;
			
			[multictrl] box {$\text{SELECT}$} k ~ j;
			hspace {5pt} -;
			measure l,j,k;
		\end{yquant}
	\end{tikzpicture}
\caption{Quantum circuit of QPS for multiple unitary matrices.}\label{figer:1}
\end{figure}

Following this approach, measuring $l$ unitary matrices would require $l$ independent circuits. By introducing idx register with $\lceil \log l\rceil$ qubits, we extend this scheme to enable measurement of all $l$ unitary matrices with just a single circuit. As shown in Fig. \ref{figer:1}, the Hamiltonian selection oracle is defined as
\begin{equation*}
	\text{SELECT} = \sum_{i=0}^{l-1} \ket{i}\bra{i} \otimes U_i.
\end{equation*}
The quantum circuit evolves the initial state $\ket{0}^{\otimes n}_{\text{w}} \ket{0}_{\text{a}}^{\otimes \lceil \log l\rceil} \ket{0}_{\text{aug}}^{\otimes n}$ as follows:
\begin{equation}
	\begin{split}
		&\qquad  \qquad \qquad\ket{0}^{\otimes n}_{\text{w}} \ket{0}_{\text{idx}}^{\otimes \lceil \log l\rceil} \ket{0}_{\text{aug}}^{\otimes n}\\
		&\xrightarrow{H^{\otimes n}\otimes H^{\lceil \log l\rceil} \otimes I^{\otimes n}}\frac{1}{\sqrt{2^{n+\lceil \log l\rceil}}}
		\sum_{i=0}^{2^n-1}
		\sum_{j=0}^{2^{\lceil \log l\rceil}-1} \ket{i}_{\text{w}}
		\ket{j}_{\text{idx}} \ket{0}_{\text{aug}}^{\otimes n}\\
		&\xrightarrow{\text{CNOT}^n}
		\frac{1}{\sqrt{2^{n+\lceil \log l\rceil}}}
		\sum_{i=0}^{2^n-1}
		\sum_{j=0}^{2^{\lceil \log l\rceil}-1}\ket{i}_{\text{w}}
		\ket{j}_{\text{idx}} \ket{i}_{\text{aug}}\\
		&\xrightarrow{\text{SELECT}}
		\frac{1}{\sqrt{2^{n+\lceil \log l\rceil}}}
		\sum_{i=0}^{2^n-1}
		\sum_{j=0}^{2^{\lceil \log l\rceil}-1}\ket{i}_{\text{w}}
		\ket{j}_{\text{idx}} U_j \ket{i}_{\text{aug}}.
	\end{split}
\end{equation}
Thus, for any $0\leqslant k\leqslant l-1$, the probability of obtaining the measurement outcome $|j\rangle_{\text{w}}|k\rangle_{\text{idx}}|i\rangle_{\text{aug}}$ is $\frac{1}{2^{n+\lceil \log l\rceil}} |\langle i|U_k|j\rangle|^{2}$.

\begin{figure}[H]
	\centering
	\includegraphics[scale=0.49]{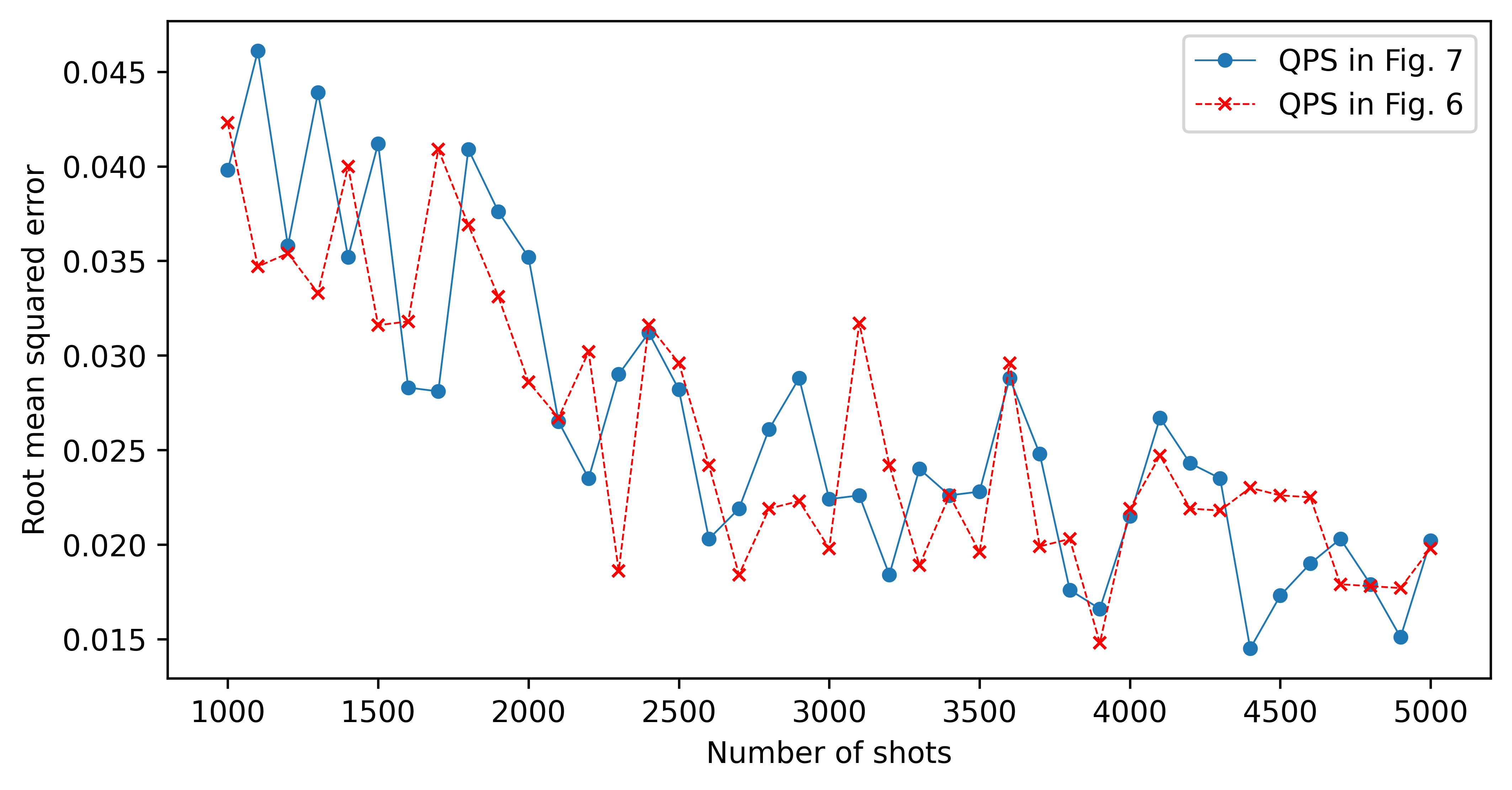}
	\caption{Experimental results of QPS for two $4\times4$ unitary matrices.}\label{QPS2}
\end{figure}

\begin{figure}[H]
	\centering
	\includegraphics[scale=0.49]{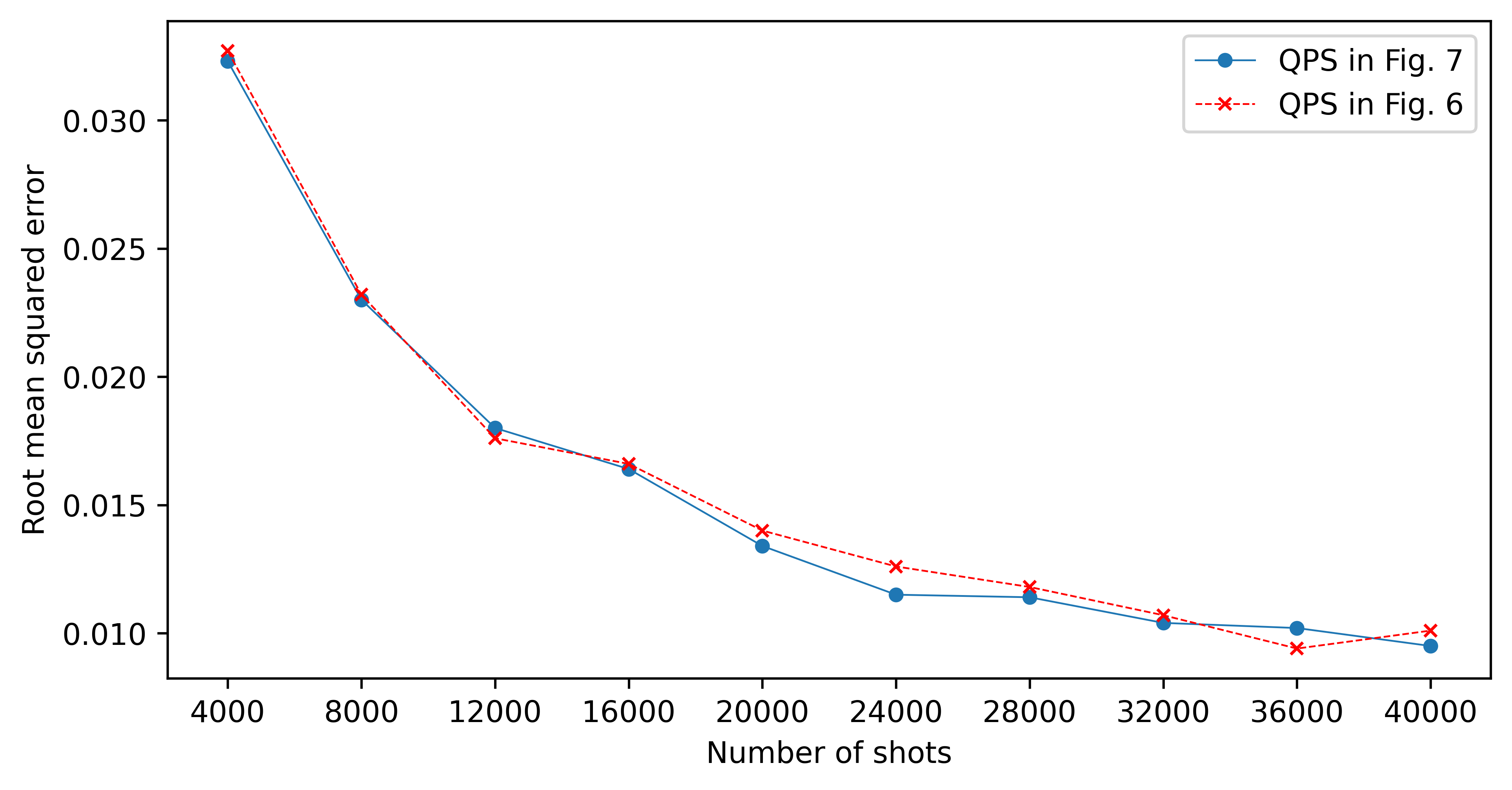}
	\caption{Experimental results of QPS for four $8\times8$ unitary matrices.}\label{QPS4}
\end{figure}

To compare the QPS circuits in Fig. \ref{QPS} with that in Fig. \ref{figer:1}, we perform numerical simulations with randomly generated unitary matrices (two $4\times4$ unitary matrices and four $8\times8$ unitary matrices). Fig. \ref{QPS2} and Fig. \ref{QPS4} show the relationship between the number of shots and the root mean squared error for both circuit designs. The results demonstrate that the circuit structure in Fig. \ref{figer:1} maintains measurement accuracy and error ranges comparable to the Fig. \ref{QPS}. For operational convenience, we adopt the scheme presented in Fig. \ref{figer:1} in this study.

\section{Data input model based on LCU}\label{section:Data}
\begin{figure}[H]
	\centering
	\begin{tikzpicture}
		\begin{yquant}
			qubit {ancilla} l;
			qubit {} j;
			["north:$n$"
			{font=\protect\footnotesize, inner sep=0pt}]
			slash j;
			["north:$m$"
			{font=\protect\footnotesize, inner sep=0pt}]
			slash l;
			hspace {5pt} -;
			box {PREP} l;
			hspace {5pt} -;
			[multictrl] box {$\text{SELECT}(A)$} j ~ l;
			hspace {5pt} -;
			box {UNPREP} l;
			hspace {5pt} -;
		\end{yquant}
	\end{tikzpicture}
	\caption{Quantum circuit of LCU~\cite{babbush2018encoding}.}
	\label{circuit: LCU}
\end{figure}
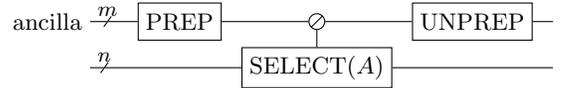

Taking matrix $A$ as an example. The LCU method needs to decompose a non-unitary matrix $A$ into a linear combination of unitary matrices
\begin{equation}
	A=\sum_{i=0}^{2^m-1}\alpha_{i}A_{i},
\end{equation}
where $A_i$ are unitaries and $\alpha_i \in \mathbb{C}$. Fig. \ref{circuit: LCU} demonstrates the corresponding quantum circuit implementation, employing state preparation oracles PREP and UNPREP to generate the quantum state,
\begin{equation*}
	\begin{split}
		\text{PREP}\ket{0}^{\otimes m}&=\text{UNPREP}^\dagger\ket{0}^{\otimes m}\\
		&=\frac{1}{\sqrt{c }}\sum_{i}\sqrt{\alpha_i}\ket{i},
	\end{split}
\end{equation*}
where the normalization constant $c=\sum\limits_{i=0}^{2^m-1}|\alpha_{i}|$, and the Hamiltonian selection oracle $\text{SELECT}(A)$ is defined as
\begin{equation*}
\begin{split}
	\text{SELECT}(A) &= \sum_{i=0}^{2^m-1} \ket{i}\bra{i}\otimes A_i,
\end{split}
\end{equation*}
which applies the unitary $A_i$ conditioned on the state $ \ket{i}$.

\begin{widetext}

\section{Derivation of the gradient formulas}\label{proof}
This appendix provides a detailed derivation of the parameter shift rule for the specific loss function $\mathcal{L}(\boldsymbol{\theta},\boldsymbol{\phi})$ defined in (\ref{loss}) of the main text. While the parameter shift rule is a known technique, using it for our problem requires a special calculation because our loss function depends on two parameterized unitary matrices, $Q^{\dag}(\boldsymbol{\theta})$ and $Z(\boldsymbol{\phi})$, and the non-unitary matrices $A$ and $B$. The following derives the explicit forms of the partial derivatives given in (\ref{derivatives}).

Consider the parameterized quantum circuit $Q(\boldsymbol{\theta})=\Pi_{i=\ell_1}^{1}Q_{i}(\theta_i)$ and $Z(\boldsymbol{\phi})=\Pi_{i=\ell_2}^{1}Z_{i}(\phi_i)$. For convenience, we define
\begin{equation}
	\begin{split}
		Q_{i:j}(\theta_{i:j})&=Q_i(\theta_i)\cdots Q_j(\theta_j)\\
		L_{ij}^{A} (\boldsymbol{\theta},\boldsymbol{\phi})&=|  \bra{i} Q^{\dag}_{1:\ell_1}(\theta_{1:\ell_1})A Z_{\ell_2:1}(\phi_{\ell_2:1})  \ket{j} |^2 \\
		L_{ij}^{B}(\boldsymbol{\theta},\boldsymbol{\phi})&= |  \bra{i} Q^{\dag}_{1:\ell_1}(\theta_{1:\ell_1})B Z_{\ell_2:1}(\phi_{\ell_2:1})  \ket{j} |^2\\
		\ket{\varphi_l}&=Q_{l-1:1}(\theta_{l-1:1})\ket{i}\\
		G=Q^{\dag}_{l+1:\ell_1}(\theta_{l+1:\ell_1})&AZ_{\ell_2:1}(\phi_{\ell_2:1})\ket{j}\bra{j}Z_{1:\ell_2}(\phi_{1:\ell_2})A^{\dag}Q_{\ell_1:l+1}(\theta_{\ell_1:l+1}).
	\end{split}
\end{equation}
The loss function can be expressed as

\begin{equation}
	\begin{split}
		\mathcal{L}(\boldsymbol{\theta},\boldsymbol{\phi})&=\sum_{i=1}^{2^n-1}\sum_{j=0}^{i-1}
		\left( |  \bra{i} Q^{\dag}_{1:\ell_1}(\theta_{1:\ell_1})A Z_{\ell_2:1}(\phi_{\ell_2:1})  \ket{j} |^2  +
		|  \bra{i} Q^{\dag}_{1:\ell_1}(\theta_{1:\ell_1})B Z_{\ell_2:1}(\phi_{\ell_2:1})  \ket{j} |^2 \right)\\
		&=\sum_{i=1}^{2^n-1}\sum_{j=0}^{i-1}\left(  L_{ij}^{A} (\boldsymbol{\theta},\boldsymbol{\phi})  +
		L_{ij}^{B} (\boldsymbol{\theta},\boldsymbol{\phi})  \right).
	\end{split}
\end{equation}
Since $Q_l(\theta_l)=\text{e}^{-i\theta_lH_l/2}$ and $Z_k(\phi_k)=\text{e}^{-i\phi_kV_k/2}$ are generated by Pauli products $H_l$ and $V_k$ respectively, we have
\begin{equation}
	\begin{split}
		\frac{\partial L_{ij}^{A} (\boldsymbol{\theta},\boldsymbol{\phi}) }{\partial \theta_l}
		&=\frac{\partial}{\partial \theta_l} \left( \bra{i} Q^{\dag}_{1:\ell_1}(\theta_{1:\ell_1})AZ_{\ell_2:1}(\phi_{\ell_2:1})\ket{j}\bra{j}Z^{\dag}_{1:\ell_2}(\phi_{1:\ell_2})A^{\dag}Q_{\ell_1:1}(\theta_{\ell_1:1})\ket{i} \right)\\
		&=\frac{\partial}{\partial \theta_l} \left(\bra{\varphi_l}Q^{\dag}_{l}(\theta_l)GQ_l(\theta_l)\ket{\varphi_l}\right)\\
		&=\bra{\varphi_l}\frac{\partial Q_{l}^{\dag}(\theta_l)}{\partial \theta_l}GQ_l(\theta_l)\ket{\varphi_l}+\bra{\varphi_l}Q^{\dag}_{l}(\theta_l)G\frac{\partial Q_{l}(\theta_l)}{\partial \theta_l}\ket{\varphi_l}\\
		&=\bra{\varphi_l} \left(  iQ_{l}^{\dag}(\theta_l) H_l/2  \right) GQ_l(\theta_l)\ket{\varphi_l}+\bra{\varphi_l}Q^{\dag}_{l}(\theta_l)G \left(  -iH_lQ_{l}(\theta_l)/2   \right) \ket{\varphi_l}\\
		&=\frac{\bra{\varphi_l} Q_{l}^{\dag}(\theta_l)  \left(  iH_lG-iGH_l  \right)  Q_{l}(\theta_l) \ket{\varphi_l} }{2}.
	\end{split}
\end{equation}
By matching the coefficients of the Taylor series, it can be shown that $Q_l(\theta_l)= \cos (\theta_l/2) \mathbb{I} -i \sin (\theta_l/2)H_l$, and
\begin{equation}
	\begin{split}
		iH_lG-iGH_l&=\frac{ (\mathbb{I} + iH_l)G(\mathbb{I}-iH_l) -(\mathbb{I}-iH_l)G(\mathbb{I}+iH_l) }{2}\\
		&=\text{e}^{i\frac{\pi}{4}H_l}G\text{e}^{-i\frac{\pi}{4}H_l} - \text{e}^{-i\frac{\pi}{4}H_l}G\text{e}^{i\frac{\pi}{4}H_l}\\
		&=Q^{\dag}_{l}(\frac{\pi}{2})GQ_l(\frac{\pi}{2})-Q^{\dag}_{l}(-\frac{\pi}{2})GQ_l(-\frac{\pi}{2}).
	\end{split}
\end{equation}
In summary, the derivative of the operator is given by
\begin{equation}
	\begin{split}
		\frac{\partial L_{ij}^{A}(\boldsymbol{\theta},\boldsymbol{\phi})}{\partial \theta_l}&=\frac{\bra{\varphi_l} Q_{l}^{\dag}(\theta_l)  \left(  Q^{\dag}_{l}(\frac{\pi}{2})GQ_l(\frac{\pi}{2})-Q^{\dag}_{l}(-\frac{\pi}{2})GQ_l(-\frac{\pi}{2})  \right)  Q_{l}(\theta_l) \ket{\varphi_l} }{2}\\
		&=\frac{\bra{\varphi_l} Q^{\dag}_{l} (\theta_{l}+\frac{\pi}{2}) G Q_{l} (\theta_{l}+\frac{\pi}{2}) \ket{\varphi_l} }{2} -
		\frac{\bra{\varphi_l} Q^{\dag}_{l} (\theta_{l}-\frac{\pi}{2}) G Q_{l} (\theta_{l}-\frac{\pi}{2}) \ket{\varphi_l} }{2}\\
		&= \frac{L_{ij}^{A} (\boldsymbol{\theta}_{l+\frac{\pi}{2}},\boldsymbol{\phi}) - L_{ij}^{A} (\boldsymbol{\theta}_{l-\frac{\pi}{2}},\boldsymbol{\phi})}{2}.
	\end{split}
\end{equation}
Similarly,
\begin{equation}
	\frac{\partial L_{ij}^{B}(\boldsymbol{\theta},\boldsymbol{\phi})}{\partial \theta_l}
	= \frac{L_{ij}^{B} (\boldsymbol{\theta}_{l+\frac{\pi}{2}},\boldsymbol{\phi}) - L_{ij}^{B} (\boldsymbol{\theta}_{l-\frac{\pi}{2}},\boldsymbol{\phi})}{2}.
\end{equation}
Thus, the gradient of the loss function is
\begin{equation}
	\begin{split}
		\frac{\partial \mathcal{L}(\boldsymbol{\theta},\boldsymbol{\phi})}{\partial \theta_l}
		&=\frac{\partial}{\partial\theta_l} \sum_{i=1}^{2^n-1}\sum_{j=0}^{i-1} \left( L^{A}_{ij}(\boldsymbol{\theta},\boldsymbol{\phi})
		+  L^{B}_{ij}(\boldsymbol{\theta},\boldsymbol{\phi}) \right)\\
		&=\sum_{i=1}^{2^n-1}\sum_{j=0}^{i-1} \left(  \frac{\partial L^{A}_{ij}(\boldsymbol{\theta},\boldsymbol{\phi}) }{\partial \theta _l} +
		\frac{\partial L^{B}_{ij}(\boldsymbol{\theta},\boldsymbol{\phi}) }{\partial \theta _l} \right)\\
		&=\sum_{i=1}^{2^n-1}\sum_{j=0}^{i-1} \left(   \frac{ L^{A}_{ij}(\boldsymbol{\theta}_{l+\frac{\pi}{2}},\boldsymbol{\phi})-  L^{A}_{ij}(\boldsymbol{\theta}_{l-\frac{\pi}{2}},\boldsymbol{\phi}) }{2}
		+ \frac{ L^{B}_{ij}(\boldsymbol{\theta}_{l+\frac{\pi}{2}},\boldsymbol{\phi})-  L^{B}_{ij}(\boldsymbol{\theta}_{l-\frac{\pi}{2}},\boldsymbol{\phi}) }{2}   \right)\\
		&= \sum_{i=1}^{2^n-1}\sum_{j=0}^{i-1} \left(   \frac{ L^{A}_{ij}(\boldsymbol{\theta}_{l+\frac{\pi}{2}},\boldsymbol{\phi})+ L^{B}_{ij}(\boldsymbol{\theta}_{l+\frac{\pi}{2}},\boldsymbol{\phi})}{2}
		- \frac{ L^{A}_{ij}(\boldsymbol{\theta}_{l-\frac{\pi}{2}},\boldsymbol{\phi})+  L^{B}_{ij}(\boldsymbol{\theta}_{l-\frac{\pi}{2}},\boldsymbol{\phi}) }{2}   \right)\\
		&=\frac{\mathcal{L}(\boldsymbol{\theta}_{l+\frac{\pi}{2}},\boldsymbol{\phi})- \mathcal{L}(\boldsymbol{\theta}_{l-\frac{\pi}{2}},\boldsymbol{\phi}) }{2}.
	\end{split}
\end{equation}

For $\frac{\partial L_{ij}^{A}(\boldsymbol{\theta},\boldsymbol{\phi})}{\partial \phi_k}$, we define
\begin{equation}
	\begin{split}
		Z_{i:j}(\phi_{i:j})&=Z_i(\phi_i)\cdots Z_j(\phi_j)\\
		\ket{\psi_k}&=Z_{k-1:1}(\phi_{k-1:1})\ket{j}\\
		W=Z^{\dag}_{k+1:\ell_2}(\phi_{k+1:\ell_2})&A^{\dag}Q_{\ell_1:1}(\theta_{\ell_1:1})\ket{i}\bra{i}Q^{\dag}_{1:\ell_1}(\theta_{1:\ell_1})AZ_{\ell_2:k+1}(\phi_{\ell_2:k+1}).
	\end{split}
\end{equation}
The derivative is computed by
\begin{equation}
	\begin{split}
		\frac{\partial L_{ij}^{A}(\boldsymbol{\theta},\boldsymbol{\phi})}{\partial\phi_k}
		&=\frac{\partial}{\partial \phi_k} \left(   \bra{j} Z^{\dag}_{1:\ell_2}(\phi_{1:\ell_2})A^{\dag}Q_{\ell_1:1} (\theta_{\ell:1}) \ket{i} \bra{i} Q^{\dag}_{1:\ell_1}(\theta_{1:\ell_1})AZ_{\ell_2:1}(\phi_{\ell_2:1}) \ket{j} \right)\\
		&=\frac{\partial}{\partial\phi_k} \left(  \bra{\psi_k} Z_k^{\dag}(\phi_k)WZ_k(\phi_k) \ket{\psi_k}  \right).
	\end{split}
\end{equation}
Since it has the same form as $\frac{\partial L_{ij}^{A}(\boldsymbol{\theta},\boldsymbol{\phi})}{\partial\theta_l}$, the result can be directly obtained as
\begin{equation}
	\begin{split}
		\frac{\partial L_{ij}^{A}(\boldsymbol{\theta},\boldsymbol{\phi})}{\partial \phi_k}&=\frac{L_{ij}^{A}(\boldsymbol{\theta},\boldsymbol{\phi}_{k+\frac{\pi}{2}})- L_{ij}^{A}(\boldsymbol{\theta},\boldsymbol{\phi}_{k-\frac{\pi}{2}}) }{2},\\
		\frac{\partial L_{ij}^{B}(\boldsymbol{\theta},\boldsymbol{\phi})}{\partial \phi_k}&=\frac{L_{ij}^{B}(\boldsymbol{\theta},\boldsymbol{\phi}_{k+\frac{\pi}{2}})- L_{ij}^{B}(\boldsymbol{\theta},\boldsymbol{\phi}_{k-\frac{\pi}{2}}) }{2},\\
		\frac{\partial \mathcal{L}(\boldsymbol{\theta},\boldsymbol{\phi})}{\partial \phi_k}&=\frac{\mathcal{L}(\boldsymbol{\theta},\boldsymbol{\phi}_{k+\frac{\pi}{2}})- \mathcal{L}(\boldsymbol{\theta},\boldsymbol{\phi}_{k-\frac{\pi}{2}}) }{2}.
		\end{split}
\end{equation}
\hfill $\square$

\end{widetext}

\section{Design of parameterized circuits}

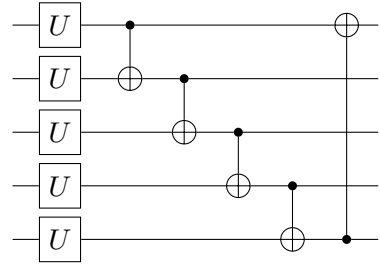
\begin{figure}[H]
	\centering
	\begin{tikzpicture}[scale=1.2]
		\begin{yquant}
			qubit {} j4; qubit {} j3; qubit {} j2; qubit {} j1; qubit {} j0;
			hspace {5pt} -;
			box {$U$} j4; box {$U$} j3; box {$U$} j2; box {$U$} j1; box {$U$} j0;
			hspace {5pt} -;
			cnot j3 | j4; cnot j2 | j3; cnot j1 | j2; cnot j0 | j1; cnot j4 | j0;
		\end{yquant}
	\end{tikzpicture}
	\caption{Schematic diagram of the parameterized quantum circuit architecture.}\label{fig:ansatz}
\end{figure}

In Section \ref{section:A} and \ref{section:C}, we employed the six-layer parameterized quantum circuit structure shown in Fig. \ref{fig:ansatz}. It should be noted that while the schematic diagram adopts a five-qubit layout for illustrative clarity, the actual parameterized circuits used in Section \ref{section:A} and \ref{section:C} are implemented with two qubits. The circuit consists of single-qubit rotation gates and two-qubit CNOT gates. When the input matrices $A$ and $B$ contained only real elements, we used $U=R_y(\theta_i)$ as the single-qubit gate. When $A$ or $B$ contained complex elements, we adopted the general rotation gate $U=R_z(\theta_{i_1})R_y(\theta_{i_2})R_z(\theta_{i_3})$. In Section \ref{section:B}, we used quantum circuit architecture search to find the best parameterized quantum circuit design for matrices $A$ and $B$.

\begin{thebibliography}{99}
\bibitem{B} W. W. Bradbury and R. Fletcher, New iterative methods for solution of the eigenproblem, Num. Math. {\bf 9}, 259 (1966).

\bibitem{MolerSIAM1973} C. B. Moler and  G. W. Stewart, An algorithm for generalized matrix eigenvalue problems, SIAM J. Numer. Anal. {\bf 10}, 241 (1973).

\bibitem{SleijpenBit1996} G. L. G. Sleijpen, A. G. L. Fokkema, D. R. Fokkema, and H. A. van der Vorst, Jacobi-davidson type methods for generalized eigenproblems and polynomial eigenproblems, Bit Numer Math {\bf 36}, 595 (1996).

\bibitem{4} F. Barahona, On the computational complexity of Ising spin glass models, J. Phys. A {\bf 15}, 3241 (1982).

\bibitem{5} S. Wiesner, Simulations of many-body quantum systems by a quantum computer, arXiv: quant-ph/9603028 (1996).

\bibitem{6} D. Poulin and P. Wocjan, Preparing ground states of quantum many-body systems on a quantum computer, Phys. Rev. Lett. {\bf 102}, 130503 (2009).

\bibitem{7} D. S. Abrams and S. Lloyd, Simulation of many-body fermi systems on a universal quantum computer, Phys. Rev. Lett. {\bf 79}, 2586 (1997).

\bibitem{8} A. Smith, M. Kim, F. Pollmann, and J. Knolle, Simulating quantum many-body dynamics on a current digital quantum computer, npj Quantum Inf. {\bf 5}, 106 (2019).

\bibitem{Aspuru-Guzik} A. Aspuru-Guzik, A. D. Dutoi, P. J. Love, and M. Head-Gordon, Simulated quantum computation of molecular energies, Science {\bf 309}, 1704 (2005).

\bibitem{Malley} P. J. O’Malley, R. Babbush, I. D. Kivlichan, J. Romero, J. R. McClean, R. Barends, J. Kelly, P. Roushan, A. Tranter, and N. Ding, Scalable quantum simulation of molecular energies, Phys. Rev. X {\bf 6}, 031007 (2016).

\bibitem{Parker} J. B. Parker and I. Joseph, Quantum phase estimation for a class of generalized eigenvalue problems, Phys. Rev. A {\bf 102}, 022422 (2020).

\bibitem{Raiche} G. Raichel-Mieldzio\'{c}, S. Pli\'{s}, and E. Zak,Quantum algorithm for solving generalized eigenvalue problems with application to the Schr\"{o}dinger equation, arXiv: quant-ph/2506.13534v1 (2025).

\bibitem{12} A. Peruzzo, J. Clean, P. Shadbolt, M. H. Yung, X. Q. Zhou,  P. J. Love, A. Aspuru-Guzik, and J. L. O'Brien, A variational eigenvalue  solver on a photonic quantum processor, Nat. Commun. {\bf 5}, 4213 (2014).

\bibitem{14} H. F. Zhao P. Zhang, and T. C Wei, A universal variational quantum eigensolver for non‑Hermitian systems, Sci. Rep. {\bf 13}, 22313 (2023).

\bibitem{15} X. D. Xie, Z. Y. Xue, and D. B. Zhang, Variational quantum algorithms for scanning the complex spectrum of non-Hermitian systems, Front. Phys. {\bf 19}, 41202 (2024).

\bibitem{16} J. M. Liang, S. Q. Shen, M Li, and S. M. Fei, Quantum algorithms for the generalized eigenvalue problem, Quantum Inf. Process. {\bf 21}, 23 (2022).

\bibitem{Sato} Y. Sato, H. C. Watanabe, R. Raymond, R. Kondo, K. Wada, K. Endo, M. Sugawara, and N. Yamamoto, Variational quantum algorithm for generalized eigenvalue problems and its application to the finite-element method, Phys. Rev. A {\bf 108}, 022429 (2023).

\bibitem{17} M. R. Hwang, E. Jung, M. Kim, and D. Park, Euclidean time method in generalized eigenvalue equation, Quantum Inf. Process. {\bf 23}, 62 (2024).

\bibitem{13} M. Cerezo, A. Arrasmith,  R. Babbush, S. C. Benjamin, S. Endo, K. Fujii, J. R. McClean, K. Mitarai, X. Yuan, and L. Cincio, Variational quantum algorithms, Nat. Rev. Phys. {\bf 3}, 625 (2021).

\bibitem{Kandala} A. Kandala, A. Mezzacapo, K. Temme, M. Takita, M. Brink, J. M. Chow, and J. M. Gambetta, Hardware-efficient variational quantum eigensolver for small molecules and quantum magnets, Nature {\bf 549}, 242 (2017).

\bibitem{Higgott} O. Higgott, D. Wang, and S. Brierley, Variational quantum computation of excited states, Quantum {\bf 3}, 156 (2019).

\bibitem{Liu} H. L. Liu, Y. S. Wu, L. C. Wan, S. J. Pan, S. J. Qin, F. Gao, and Q. Y. Wen, Variational quantum algorithm for the poisson equation, Phys. Rev. A {\bf 104}, 022418 (2021).

\bibitem{Singular} X. Wang, Z. X. Song, and Y. L. Wang, Variational quantum singular value decomposition, Quantum {\bf 5}, 483 (2021).

\bibitem{G} G. H. Golub and C. F. V. Loan, $Matrix$ $Computations$, 4th. ed. (Johns Hopkins University Press, Baltimore, 2013).

\bibitem{37} A. Gily\'{e}n, Y. Su, G. H. Low, and N. Quantum singular value transformation and beyond: exponential improvements for quantum matrix arithmetics, in $Proc$. $51st$ $Annual$ $ACM$ $SIGACT$ $Symposium$ $on$ $Theory$ $of$ $Computing$ (ACM, 2019) pp. 193–204.

\bibitem{Yang} C. L. Yang, Z. X. Li, H. M. Yao, Z. B. Fan, G. F. Zhang, and J. S. Liu, Dictionary-based sparse block encoding with low subnormalization and circuit depth, Quantum {\bf 9}, 1805 (2025).

\bibitem{LZX} Z. X. Li, X. M. Zhang, C. L. Yang, and G. F. Zhang, Binary tree block encoding of classical matrix, arXiv: quant-ph/2504.05624 (2025).

\bibitem{38} G. L. Long, General quantum interference principle and duality computer, C. Theor. Phys. {\bf 45}, 825 (2006).

\bibitem{Childs} A. M. Childs and N. Wiebe, Hamiltonian simulation using linear combinations of unitary operations, Quant. Inf. Comput. {\bf 12}, 901 (2012).

\bibitem{35} M. Cerezo, K. Sharma, A. Arrasmith, and P. J. Coles, Variational quantum state eigensolver, npj Quantum Inf. {\bf 8}, 113 (2022).

\bibitem{Mitarai} K. Mitarai, M. Negoro, M. Kitagawa, and K. Fujii, Quantum circuit learning, Phys. Rev. A {\bf 98}, 032309 (2018).

\bibitem{Nocedal} J. Nocedal and S. J. Wright, $Numerical$ $Optimization$, (Springer, New York, 2006).

\bibitem{Jarrod} J. R. McClean, S. Boixo, V. N. Smelyanskiy, R. Babbush, and H. Neven, Barren plateaus in quantum neural network training landscapes, Nat. Commun. {\bf 9}, 4812 (2018).

\bibitem{Edward} E. Grant, L. Wossnig, M. Ostaszewski, and M. Benedetti, An initialization strategy for addressing barren plateaus in parametrized quantum circuits, Quantum {\bf 3}, 241 (2019).

\bibitem{NatCommun12} M. Cerezo, A. Sone, T. Volkoff, L. Cincio, and P. J. Coles, Cost function dependent barren plateaus in shallow parametrized quantum circuits, Nat. Commun. {\bf 12}, 1791 (2021).

\bibitem{36} OriginQ, Pyqpanda, \url{https://qcloud.originqc.com.cn}.

\bibitem{40} \url {https://github.com/LiJiaxin888/VQGE}.

\bibitem{jensen} F. B. Jensen, W. A. Kuperman, M. B. Porter, H. Schmidt, and A. Tolstoy, $Computational$ $Ocean$ $Acoustics$, (Springer, New York, 2011).

\bibitem{aki} K. Aki and P. G. Richards, $Quantitative$ $Seismology$, 2nd ed. (University Science Books, Sausalito, California, 2002).

\bibitem{Errormitigation} A. Kandala, K. Temme, A. D. C\'{o}rcoles, A. Mezzacapo, J. M. Chow, and J. M. Gambetta, Error mitigation extends the computational reach of a noisy quantum processor, Nature {\bf 567}, 491 (2019).

\bibitem{TemmeErrormitigation2017} K. Temme, S. Bravyi, and J. M. Gambetta, Error mitigation for short-depth quantum circuits, Phys. Rev. Lett. {\bf 119}, 180509 (2017).

\bibitem{EndoPractical2018} S. Endo, S. C. Benjamin, and Y. Li, Practical quantum error mitigation for near-future applications, Phys. Rev. X {\bf 8}, 031027 (2018).

\bibitem{CzarnikError2021} P. Czarnik, A. Arrasmith, P. J. Coles, and L. Cincio, Error mitigation with Clifford quantum-circuit data, Quantum {\bf 5}, 592 (2021).

\bibitem{McCleanHybrid2017} J. R. McClean, M. E. Kimchi-Schwartz, J. Carter, and W. A. de Jong, Hybrid quantum-classical hierarchy for mitigation of decoherence and determination of excited states, Phys. Rev. A {\bf 95}, 042308 (2017).

\bibitem{babbush2018encoding} R. Babbush, C. Gidney, D. W. Berry, N. Wiebe, J. McClean, A. Paler, A. Fowler, and H. Neven, Encoding electronic spectra in quantum circuits with linear t complexity, Phys. Rev. X {\bf 8}, 041015 (2018).


\end{thebibliography}


\setcounter{figure}{0}
\setcounter{section}{0}
\setcounter{equation}{0}
\renewcommand{\theequation}{S\arabic{equation}}
\renewcommand{\thefigure}{S\arabic{figure}}


\end{document}